\documentclass{article}
\usepackage{waspaa23,amsmath,graphicx,url,times}
\usepackage{color}

\usepackage{amssymb}
\usepackage{todonotes}
\usepackage{stmaryrd}
\usepackage[utf8]{inputenc}
\usepackage{hyperref}

\usepackage{dsfont}
\usepackage{caption}
\usepackage{subcaption}
\usepackage{tikz}
\usepackage{csquotes}
\usepackage{multirow}

\newcommand{\BoundSet}{Z}
\newcommand{\aBound}{\zeta}
\newcommand{\aSeg}{S}
\newcommand{\etal}{\textit{et al.}}
\newcommand{\ie}{\textit{i.e.} }
\newcommand{\eg}{\textit{e.g.} }

\DeclareMathOperator*{\Fzf}{F_{0.5s}}

\DeclareMathOperator*{\Fth}{F_{3s}}

\newcommand{\SegScoreNotation}{u}

\DeclareMathOperator*{\argmax}{arg\,max}

\newcommand{\drawmatr}[6]{
    \draw[black,fill=#6] (#1,#2,#3) -- ++(#4,0,0) -- ++(0,-#5,0) -- ++(-#4,0,0) -- cycle;
}

\title{Convolutive Block-Matching Segmentation Algorithm \\with Application to Music Structure Analysis}

\name{Axel Marmoret$^\ast$,$^{1,2}$
 J\'er\'emy E. Cohen\sthanks{With the support of ANR JCJC LoRAiA ANR-20-CE23-0010.},$^{3}$
 Fr\'ed\'eric Bimbot$^{1}$}
\address{$^1$ Univ Rennes, Inria, CNRS, IRISA, France\\
$^2$ IMT Atlantique, Lab-STICC, UMR CNRS 6285, F-29238 Brest, France\\
$^3$ CNRS, CREATIS, Villeurbanne France\\
}

\begin{document}

\ninept
\maketitle

\begin{sloppy}

\begin{abstract}
  Music Structure Analysis (MSA) consists of representing a song in sections (such as ``chorus'', ``verse'', ``solo'' etc), and can be seen as the retrieval of a simplified organization of the song. This work presents a new algorithm, called Convolutive Block-Matching (CBM) algorithm, devoted to MSA. In particular, the CBM algorithm is a dynamic programming algorithm, applying on autosimilarity matrices, a standard tool in MSA. In this work, autosimilarity matrices are computed from the feature representation of an audio signal, and time is sampled on the barscale. We study three different similarity functions for the computation of autosimilarity matrices. We report that the proposed algorithm achieves a level of performance competitive to that of supervised State-of-the-Art methods on 3 among 4 metrics, while being unsupervised.
\end{abstract}


\section{Introduction}
Citing Paulus \etal~\cite{paulus2010state}, ``[...] it is the structure, or the relationships between the sound events that create musical meaning''. Following that statement, the Music Structure Analysis task (MSA) was developed, focusing on the retrieval of the structure in a song~\cite{nieto2020segmentationreview}, which can be seen as retrieving a simplified description of the organization of a song at the macroscopic scale. In particular, this work focuses on a ``flat'' estimation, \ie a single-level estimation, as opposed to a hierarchical level of structure. 
This paper specifically studies the subtask of structural segmentation of audio signals, consisting of estimating the boundaries between different sections, \ie estimating the time instances which separate consecutive sections on the basis of an audio recording. 


This task is generally solved focusing on one or several criteria among four: homogeneity, novelty, repetition and regularity~\cite{nieto2020segmentationreview}. 
The homogeneity criterion assumes that musical elements (notes, chords, tonality, timbre, ...) should be similar to constitute a section. Novelty is the counterpart of homogeneity, considering that boundaries must be placed between consecutive musical elements that are highly contrastive. 
Both are generally studied jointly. 
Repetition does not consider segments locally, but rather relies on a global approach of the song, to catch recurring motifs (for instance a melodic line). The rationale is that motifs, which may be individually heterogeneous, constitute segments when they are repeated. Finally, the regularity criterion assumes that, within a song, segments should be of comparable sizes. 


In this article, we introduce a new algorithm to perform unsupervised music segmentation. Our contribution is the design of a (customizable) cost function which relies on homogeneity and regularity, which can be optimized using dynamic programming. This results in an efficient unsupervised segmentation technique\footnote{We acknowledge that the use of a learning-based toolbox for the bar estimation, as well as annotated examples to fit internal hyperparameters and draw hypotheses, constitute a form of supervision, that can be qualified as ``weak supervision''.}, as supported by a comparison with State-of-the-Art unsupervised and supervised methods. The algorithm is implemented in the open-source \textit{as\_seg} toolbox~\cite{asSeg}. 

This article is organized as follows: Section~\ref{sec:autosimilarity} presents the autosimilarity matrix, on which the CBM algorithm detailed in Section~\ref{sec:cbm} is applied, and Section~\ref{sec:experiments_cbm} presents experimental results, conducted on both RWC Pop and SALAMI datasets~\cite{rwc, salami}.

\subsection{Related work}
Music structure is frequently estimated using an ``autosimilarity matrix'', presented in Section~\ref{sec:autosimilarity}. In a nutshell, an autosimilarity matrix is a square matrix presenting the similarity between each pair of time instances in the song. For instance, the kernel of Foote~\cite{foote2000automatic} 
uses autosimilarity matrices, and exploits the homogeneity/novelty criteria, estimating boundaries as points of high dissimilarity between the recent past and the near future. 

Jensen~\cite{jensen2006multiple} developed an optimization problem minimizing the average self-dissimilarity (in an autosimilarity matrix) of each segment, as a way to account for the homogeneity of each segment. This optimization problem is solved by dynamic programming. Sargent \etal~\cite{sargent2016estimating} later extended this optimization paradigm by adding prior knowledge on the segment sizes in the form of constraints.

McFee \& Ellis~\cite{mcfee2014analyzing} developed an algorithm based on spectral clustering, interpreting the repetition of musical content as principally connected vertices in a graph. The structure is then obtained by studying the eigenvectors of the Laplacian of this graph, forming cluster classes for segmentation. 

Serr\`a \etal~\cite{serra2014unsupervised} developed ``Structural Features'', which, by design, encode both repetitive and homogeneous parts. The rationale of these features is to account for the musical content of several consecutive frames, then used to compute the similarity. 
Boundaries are obtained as points of high novelty between consecutive structural features.


The most recent techniques often make use of neural networks, \eg~\cite{grill2015cnn,mccallum2019unsupervised, wang2021supervised, salamon2021deep}. In particular, the work of Grill \& Schl\"uter~\cite{grill2015cnn}, based on a Convolutional Neural Network (CNN) which directly outputs estimated boundaries, may be considered as State-of-the-Art in the structural segmentation task. On the other hand, the other neural network-based techniques~\cite{mccallum2019unsupervised, wang2021supervised, salamon2021deep} develop deep embeddings on which are computed autosimilarity matrices, then post-processed into estimated boundaries (using either the algorithm of Foote~\cite{foote2000automatic} or of McFee \& Ellis~\cite{mcfee2014analyzing}).

\section{Autosimilarity Matrix}\label{sec:autosimilarity}

\subsection{Barwise TF matrix}
We consider that bars are well suited to express patterns and sections in Western modern music (our case study). In particular, we assume that musical sections start and end on downbeats, which is experimentally confirmed by works such as~\cite{mauch2009using, fuentes2019music} where the use of structural information improves the estimation of downbeats. The direct consequence is the need for a powerful tool to estimate bars, in our case the \textit{madmom} toolbox~\cite{madmom}.

Thus, and following our previous work~\cite{marmoret2022barwise}, we decide to represent music as barwise spectrograms, with a fixed number of time frames $T$ per bar, set to $T = 96$. In practice, this article focuses on the Barwise TF matrix defined in~\cite{marmoret2022barwise}, consisting of a matrix of size $B \times TF$, $B$ being the number of bars in the song (\ie a dimension accounting for the barscale), and $TF$ the vectorization of both time (at barscale) and frequency dimensions into a 
Time-Frequency dimension. Following the work of Grill \& Schl\"uter~\cite{grill2015cnn}, in what follows the frequency dimension corresponds to the Log Mel feature (\ie the logarithm of Mel coefficients) with $F = 80$, and are computed using the \textit{librosa} toolbox~\cite{mcfee2015librosa}. 

\subsection{Similarity function}
Given a Barwise TF matrix $X \in \mathbb{R}^{B\times TF}$, an autosimilarity of $X$ is defined as a matrix $A(X) \in \mathbb{R}^{B\times B}$ where each coefficient $A(X)_{ij}$ represents the similarity between two bars $X_i$ and $X_j \in \mathbb{R}^{TF}$, such that $A(X)_{ij} = s(X_i, X_j)$, subject to a similarity function $s()$. Three similarity functions are studied:

\begin{itemize}
    \item[-] \underline{Cosine similarity}, the normalized dot products between two vectors: $s^{cos}(X_i, X_j) = \frac{\left\langle X_i, X_j\right\rangle}{\|X_i\|_2 \|X_j\|_2}$.
    
    \item[-] \underline{Covariance similarity}, the Cosine similarity of the centered matrix $X$, \ie denoting as $\bar{x} \in \mathbb{R}^{TF}$ the average of all bars in the song:
    $s^{cov}(X_i, X_j) = \frac{\left\langle X_i - \bar{x}, X_j - \bar{x} \right\rangle}{\|X_i - \bar{x}\|_2 \|X_j - \bar{x}\|_2}$.
    
    \item[-] \underline{The Radial Basis Function (RBF)}, a kernel function where $s^{RBF}(X_i, X_j) = \exp\left(-\gamma\left\|\frac{X_i}{\|X_i\|_2} - \frac{X_j}{\|X_j\|_2}\right\|^2_2\right)$.
    
    Parameter $\gamma$ is set relatively to the standard deviation of the pairwise Euclidean distances, \ie denoting as $\sigma~=~\underset{1 < i,j < B, i \neq j}{\text{std}}\left(\left\|\frac{X_i}{\|X_i\|_2} - \frac{X_j}{\|X_j\|_2}\right\|^2_2 \right)$, we set $\gamma~=~\frac{1}{2\sigma}$.
\end{itemize}

These three autosimilarities are presented in Fig.~\ref{fig:different_autosimilarities}. Note that the self-similarity of a vector is equal to one in these similarity functions, and corresponds to the main diagonal of the autosimilarity matrix. 

\begin{figure}[t!]
\centering
  \includegraphics[width=\columnwidth]{ 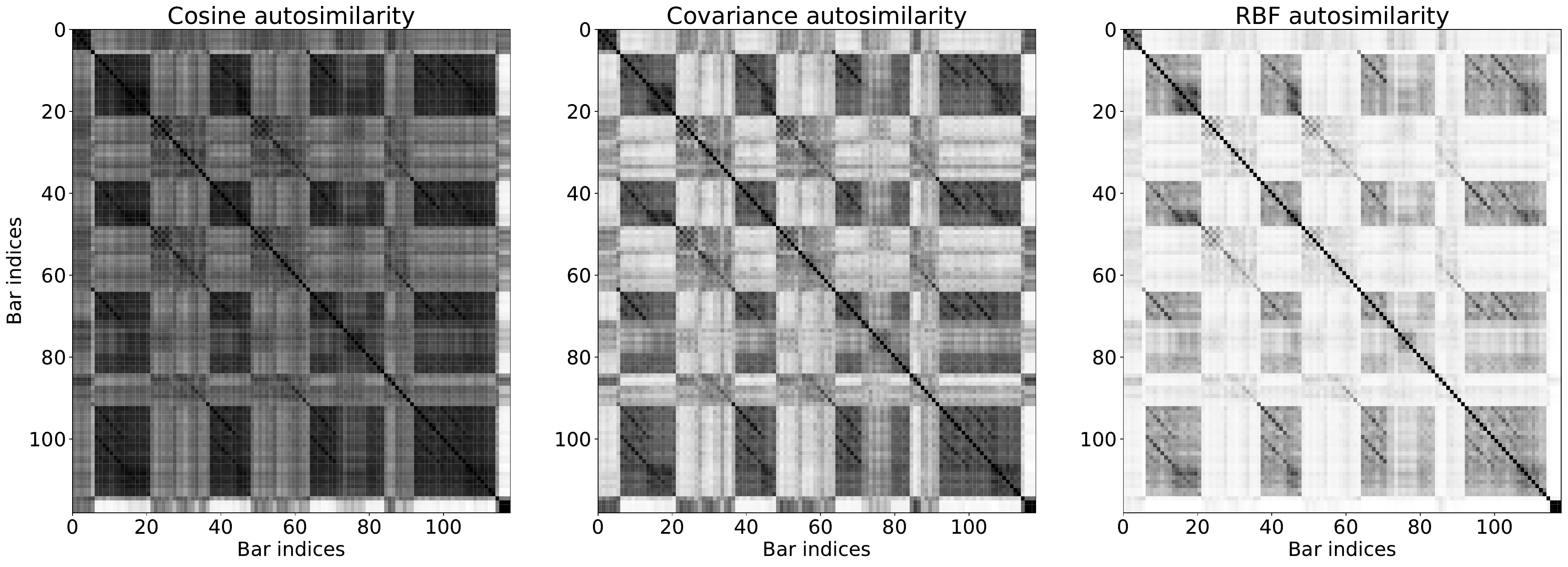}
 \caption{Cosine, Covariance and RBF autosimilarities on the song \textit{POP01} from RWC Pop~\cite{rwc}.}
  \vspace{-10pt}
\label{fig:different_autosimilarities}
\end{figure}

\section{Convolutive ``Block-Matching'' Algorithm}\label{sec:cbm}
Formally, given a musical song sampled in time as $B$ bars, structural segmentation can be defined as finding a set of boundaries (located on downbeats) $\BoundSet$ representing the start of all segments, \ie $\BoundSet~=~\{\aBound_i\in \llbracket 1, B \rrbracket , \, i \in \llbracket 1, E \rrbracket\}$, with $E \leq B$ the number of estimated boundaries.
The $i$-th segment $\aSeg_i$ is exactly the interval composed of the bars between two consecutive boundaries, \ie $\aSeg_i~=~\llbracket \aBound_i, \aBound_{i+1}\llbracket$. 

Starting from an autosimilarity matrix, structural segmentation is obtained using the Convolutive ``Block-Matching'' segmentation algorithm (CBM). In a nutshell, the CBM algorithm is based on the definition of a score function $\SegScoreNotation$ applied on segments, and the segmentation of the song results in the maximum total score of the segments, \ie denoting as $\Theta$ the set of all possible segmentations:
\begin{equation}
\BoundSet^* = \underset{\BoundSet \in \Theta}{\argmax} \sum\limits_{i = 1}^{E-1} \SegScoreNotation(\llbracket \aBound_i, \aBound_{i+1}\llbracket).
\label{eq:optimization_problem}
\end{equation}

\subsection{Problem modeling}
The optimization problem defined in (\ref{eq:optimization_problem}) can be solved using dynamic programming~\cite{bellman1952theory}. In particular, following~\cite{jensen2006multiple}, by considering each bar as a vertex and each segment between two bars as an edge, a segmentation can be reinterpreted as a path in a Directed Acyclic Graph. By assigning the segment score as the length of the associated edge, the optimal segmentation can be reframed as the problem of finding the longest path in the graph. Precisely, the solution is computed as presented in the ``Single-Source Shortest Paths in Directed Acyclic Graphs'' problem (for a critical path)~\cite[Chap. 24]{cormen2009introduction}, as the graph is topologically sorted (due to the chronological order of bars in the song), and composed of a single vertex as origin (the first bar of the song).

\subsection{Score function}
The score function in the CBM algorithm, defining the optimization problem to be solved, is inspired from the work of Sargent \etal~\cite{sargent2016estimating}, which extended the score function of Jensen~\cite{jensen2006multiple} in order to take into account both the homogeneity and the regularity criteria. For a segment $S_i$ of size $n$, this results in a mixed score function:
\begin{equation}
\label{eq:mixed_function}
  \SegScoreNotation(S_i) = \SegScoreNotation^{K}(S_i) - \lambda p(n).
\end{equation} 
In details, (\ref{eq:mixed_function}) is the weighted sum of two terms: $\SegScoreNotation^{K}(S_i)$, evaluating the homogeneity of the segment using convolution kernels (Section~\ref{sec:convolution_kernels}), and $p(n)$, a penalization term related to the regularity of the segment (Section~\ref{sec:penalty_function}). 
Parameter $\lambda$ is a weighting parameter, fitted during the experiments. 




\subsubsection{Convolution kernels}
\label{sec:convolution_kernels}
Given an autosimilarity matrix $A(X)$, the convolution score $\SegScoreNotation^K(S_i)$ of segment $\aSeg_i$ is computed by evaluating the autosimilarity values restricted to that segment, denoted as $A_{S_i}(X)$, and represents (to some extent) the inner similarity of this segment. In practice, this is obtained by weighting the different values in the autosimilarity, in a convolution operation between the autosimilarity and a (fixed) ``convolution'' kernel matrix $K$ of the size of the segment, such as:
\begin{equation}
  \SegScoreNotation^K(S_i) = \frac{1}{\nu \, n} \sum\limits_{k = 1}^{n} \sum\limits_{l = 1}^{n} A_{S_i}(X)_{kl} K_{kl}.
\end{equation}

The convolution is normalized by the size of the segment $n$, and by a parameter $\nu$, scaled on the convolution values in this particular song. Parameter $\nu$ is set as the maximal convolution value obtained by sliding a kernel of size 8 on this autosimilarity, \ie the highest score among all possible segments of size 8.


In the CBM algorithm, the design of the convolution kernel defines how to transform bar similarities into segment homogeneity, which is of particular importance. A very simple kernel is a kernel matrix full of ones, \ie $K = \mathds{1}_{n\times n}$, resulting in the sum of all the values in the autosimilarity. Still, we consider that the main diagonal in the autosimilarity is not informative regarding the overall similarity in the segment, as its values are normalized to one, thus $K_{ii} = 0, \, \forall i$. Hereafter are presented two ways for designing kernels. 

  \textbf{Full kernel}: The first kernel is called the full kernel, and corresponds to a matrix full of 1 (except on the diagonal where it is equal to 0). The full kernel captures the average value of similarities in this segment (without the self-similarity values). Practically, denoting as $K^{f}$ the full kernel, $K^{f}_{ij} =
\begin{cases} 1 & \text{if } i \neq j \\
0 & \text{if } i = j
\end{cases}.$

\textbf{Band kernel}: The second kernel design, called band kernel, emphasizes on short-term similarity: the score is computed on a few bars in the segment only, depending on their temporal proximity (only the closest bars are considered). In practice, this is obtained by setting every entry to 0, except for some upper- and sub-diagonals, where they are set to 1. The number of upper- and sub-diagonals is a parameter, corresponding to the maximal number of neighbouring bars considered to evaluate the similarity. Denoting as $v$ the number of bands, the $v$-band kernel $K^{vb}$ is defined as: $\begin{array}{crl}
K^{vb}_{ij} = \begin{cases} 1 & \text{if } 1 \leq |i - j| \leq v \\ 0 & \text{if } i = j \text{ or } |i - j| > v \end{cases}.
\end{array}$

Fig.~\ref{fig:bands_kernel} presents the full, 3-band and 7-band kernels of size 10.

\begin{figure}[t!]
  \centering
  \begin{subfigure}{0.3\columnwidth}
  \centering

    \begin{tikzpicture}[scale=2.05]
      \drawmatr{0}{0}{0}{1}{1}{black}
      \drawmatr{0}{-0.1}{0}{0.1}{0.4}{black}
      \drawmatr{0}{0}{0}{0.1}{0.1}{white}
      \drawmatr{0.1}{-0.1}{0}{0.1}{0.1}{white}
      \drawmatr{0.2}{-0.2}{0}{0.1}{0.1}{white}
      \drawmatr{0.3}{-0.3}{0}{0.1}{0.1}{white}
      \drawmatr{0.4}{-0.4}{0}{0.1}{0.1}{white}
      \drawmatr{0.5}{-0.5}{0}{0.1}{0.1}{white}
      \drawmatr{0.6}{-0.6}{0}{0.1}{0.1}{white}
      \drawmatr{0.7}{-0.7}{0}{0.1}{0.1}{white}
      \drawmatr{0.8}{-0.8}{0}{0.1}{0.1}{white}
      \drawmatr{0.9}{-0.9}{0}{0.1}{0.1}{white}

\end{tikzpicture}
    \caption{Full kernel.}
  \end{subfigure}
  \,
  \begin{subfigure}{0.3\columnwidth}
    \centering

    \begin{tikzpicture}[scale=2.05]
    \drawmatr{0}{0}{0}{1}{1}{white}

    \drawmatr{0}{-0.1}{0}{0.1}{0.3}{black}
    \drawmatr{0.1}{-0.2}{0}{0.1}{0.3}{black}
    \drawmatr{0.2}{-0.3}{0}{0.1}{0.3}{black}
    \drawmatr{0.3}{-0.4}{0}{0.1}{0.3}{black}
    \drawmatr{0.4}{-0.5}{0}{0.1}{0.3}{black}
    \drawmatr{0.5}{-0.6}{0}{0.1}{0.3}{black}
    \drawmatr{0.6}{-0.7}{0}{0.1}{0.3}{black}
    \drawmatr{0.7}{-0.8}{0}{0.1}{0.2}{black}
    \drawmatr{0.8}{-0.9}{0}{0.1}{0.1}{black}

    \drawmatr{0.1}{0}{0}{0.3}{0.1}{black}
    \drawmatr{0.2}{-0.1}{0}{0.3}{0.1}{black}
    \drawmatr{0.3}{-0.2}{0}{0.3}{0.1}{black}
    \drawmatr{0.4}{-0.3}{0}{0.3}{0.1}{black}
    \drawmatr{0.5}{-0.4}{0}{0.3}{0.1}{black}
    \drawmatr{0.6}{-0.5}{0}{0.3}{0.1}{black}
    \drawmatr{0.7}{-0.6}{0}{0.3}{0.1}{black}
    \drawmatr{0.8}{-0.7}{0}{0.2}{0.1}{black}
    \drawmatr{0.9}{-0.8}{0}{0.1}{0.1}{black}

    \drawmatr{0}{-0.5}{0}{0.1}{0.1}{white}
    \drawmatr{0.1}{-0.6}{0}{0.1}{0.1}{white}
    \drawmatr{0.2}{-0.7}{0}{0.1}{0.1}{white}
    \drawmatr{0.3}{-0.8}{0}{0.1}{0.1}{white}
    \drawmatr{0.4}{-0.9}{0}{0.1}{0.1}{white}

    \drawmatr{0.5}{0}{0}{0.1}{0.1}{white}
    \drawmatr{0.6}{-0.1}{0}{0.1}{0.1}{white}
    \drawmatr{0.7}{-0.2}{0}{0.1}{0.1}{white}
    \drawmatr{0.8}{-0.3}{0}{0.1}{0.1}{white}
    \drawmatr{0.9}{-0.4}{0}{0.1}{0.1}{white}

    \drawmatr{0}{-0.7}{0}{0.1}{0.1}{white}
    \drawmatr{0.1}{-0.8}{0}{0.1}{0.1}{white}
    \drawmatr{0.2}{-0.9}{0}{0.1}{0.1}{white}

    \drawmatr{0.7}{0}{0}{0.1}{0.1}{white}
    \drawmatr{0.8}{-0.1}{0}{0.1}{0.1}{white}
    \drawmatr{0.9}{-0.2}{0}{0.1}{0.1}{white}

    \drawmatr{0}{-0.9}{0}{0.1}{0.1}{white}

    \drawmatr{0.9}{0}{0}{0.1}{0.1}{white}

  \end{tikzpicture}
    \caption{3-band kernel.}
    \end{subfigure}
    \,
    \begin{subfigure}{0.3\columnwidth}
      \centering

      \begin{tikzpicture}[scale=2.05]
      \drawmatr{0}{0}{0}{1}{1}{white}

      \drawmatr{0}{-0.1}{0}{0.1}{0.7}{black}
      \drawmatr{0.1}{-0.2}{0}{0.1}{0.7}{black}
      \drawmatr{0.2}{-0.3}{0}{0.1}{0.7}{black}
      \drawmatr{0.3}{-0.4}{0}{0.1}{0.6}{black}
      \drawmatr{0.4}{-0.5}{0}{0.1}{0.5}{black}
      \drawmatr{0.5}{-0.6}{0}{0.1}{0.4}{black}
      \drawmatr{0.6}{-0.7}{0}{0.1}{0.3}{black}
      \drawmatr{0.7}{-0.8}{0}{0.1}{0.2}{black}
      \drawmatr{0.8}{-0.9}{0}{0.1}{0.1}{black}

      \drawmatr{0.1}{0}{0}{0.7}{0.1}{black}
      \drawmatr{0.2}{-0.1}{0}{0.7}{0.1}{black}
      \drawmatr{0.3}{-0.2}{0}{0.7}{0.1}{black}
      \drawmatr{0.4}{-0.3}{0}{0.6}{0.1}{black}
      \drawmatr{0.5}{-0.4}{0}{0.5}{0.1}{black}
      \drawmatr{0.6}{-0.5}{0}{0.4}{0.1}{black}
      \drawmatr{0.7}{-0.6}{0}{0.3}{0.1}{black}
      \drawmatr{0.8}{-0.7}{0}{0.2}{0.1}{black}
      \drawmatr{0.9}{-0.8}{0}{0.1}{0.1}{black}

      \drawmatr{0}{-0.9}{0}{0.1}{0.1}{white}

      \drawmatr{0.9}{0}{0}{0.1}{0.1}{white}

      \drawmatr{1.1}{-0.4}{0}{0.1}{0.1}{black}
      \node at (1.3,-0.45) {1};
      \drawmatr{1.1}{-0.6}{0}{0.1}{0.1}{white}
      \node at (1.3,-0.65) {0};
    \end{tikzpicture}

      \caption{7-band kernel.}
      \end{subfigure}
       \vspace{-5pt}
  \caption{Different kernels, of size 10}
  \vspace{-10pt}
  \label{fig:bands_kernel}
\end{figure}
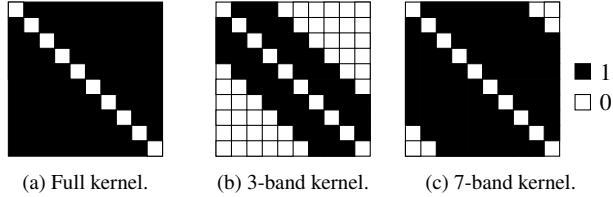

\subsubsection{Penalty function}
\label{sec:penalty_function}
The regularity function is based on prior knowledge, and aims at enforcing particular sizes of segments, which are known to be common in a number of music genres, notably Western modern music. In particular, as presented in~\cite{sargent2016estimating, marmoret2022barwise}, some sizes of segments are most common in the annotations. In particular, in the SALAMI dataset~\cite{salami}, as presented in Fig.~\ref{fig:distrib_barwise_segment_sizes_cbm_annotation}, most segments are of size 8, and the remaining segments are generally of size 4, 12 or 16. Finally,
even segments are more common than segments of odd sizes. Hence, the regularity function models this distribution, as: 
\begin{equation}
 \vspace{-5pt}
p(n) = \begin{cases}
  0 & \text{if } n = 8 \\
  \frac{1}{4} & \text{else if } n \equiv 0 \pmod 4 \\
  \frac{1}{2} & \text{else if } n \equiv 0 \pmod 2 \\
   1 & \text{otherwise}
 \end{cases}
  \vspace{-10pt}
 \end{equation}
 
\begin{figure}[t!]
 \vspace{-5pt}
\centering
\includegraphics[width=0.75\columnwidth]{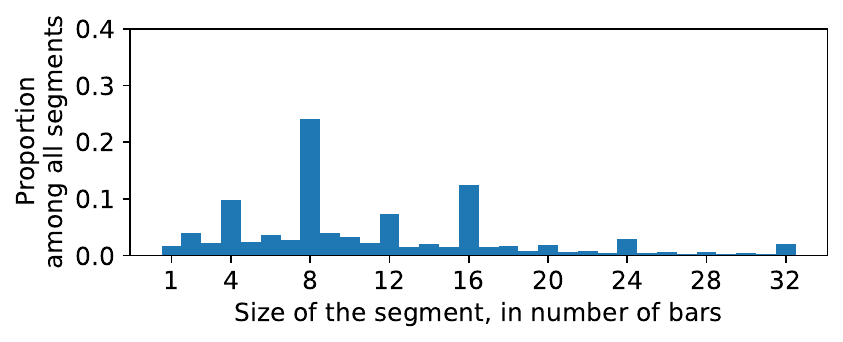}
 \vspace{-10pt}
\caption{SALAMI annotations, at bar level (coarse level, test subset defined in~\cite{grill2015cnn}).}
  \vspace{-10pt}
\label{fig:distrib_barwise_segment_sizes_cbm_annotation}
\end{figure}

\section{Experiments}
\label{sec:experiments_cbm}
\begin{figure*}[t!]
 \vspace{-10pt}
\centering
\begin{subfigure}{0.9\columnwidth}
  \includegraphics[width=\columnwidth]{ 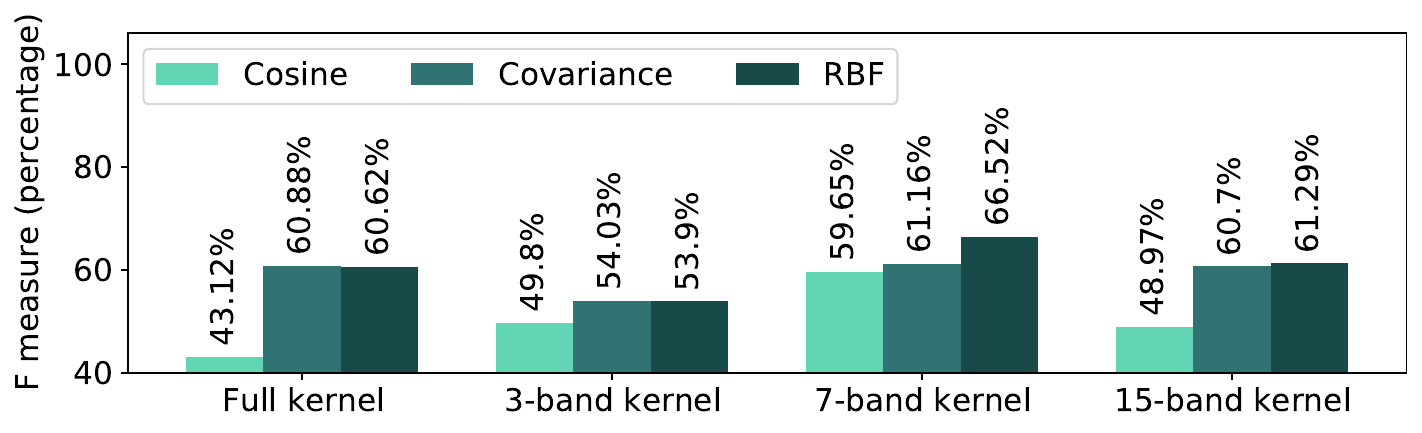}
  \caption{$\Fzf$.}
\end{subfigure}
\qquad
\begin{subfigure}{0.9\columnwidth}
  \includegraphics[width=\columnwidth]{ 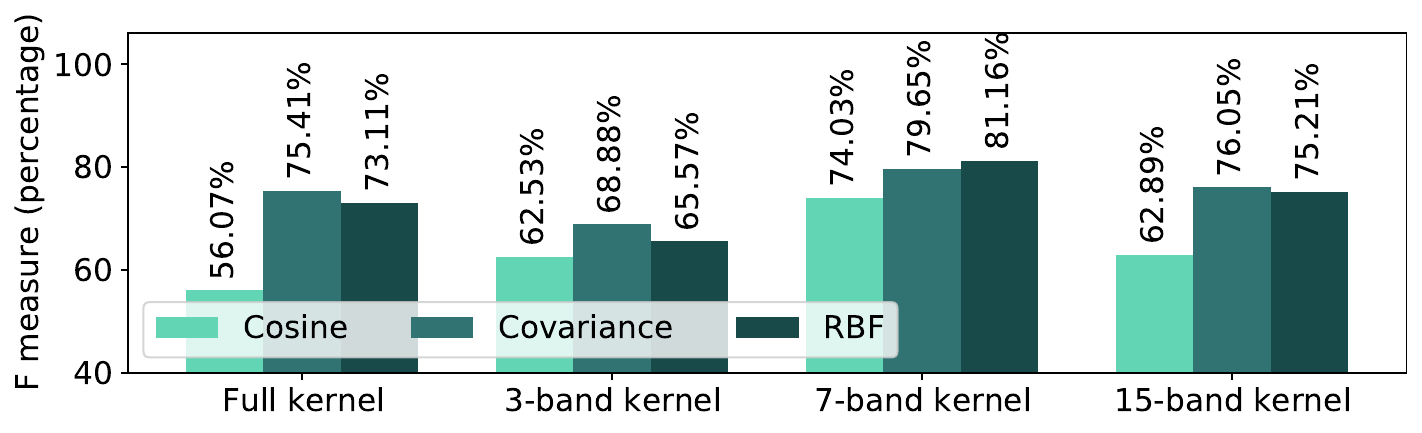}
  \caption{$\Fth$.}
  \end{subfigure}
   \vspace{-8pt}
\caption{F-measures according to the different autosimilarity matrices and kernels (RWC Pop dataset).}
 \vspace{-5pt}
\label{fig:BTF_bands_rwc}
\end{figure*}

\begin{figure*}[t!]
\centering
\begin{subfigure}{0.9\columnwidth}
     \vspace{-4pt}
  \includegraphics[width=\columnwidth]{ 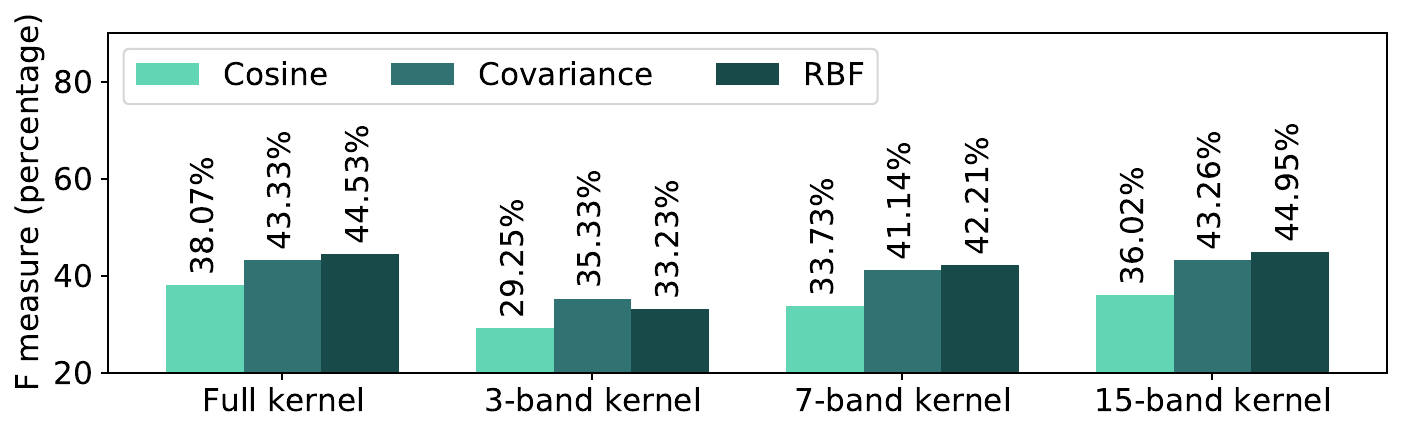}
  \caption{$\Fzf$.}
\end{subfigure}
\qquad
\begin{subfigure}{0.9\columnwidth}
     \vspace{-4pt}
  \includegraphics[width=\columnwidth]{ 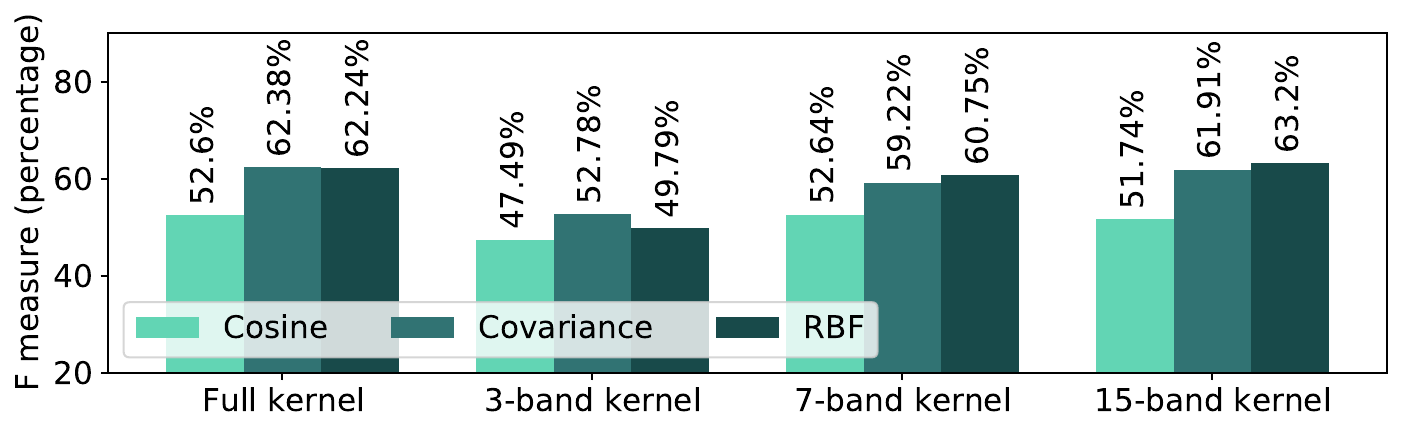}
  \caption{$\Fth$.}
  \end{subfigure}
   \vspace{-8pt}
\caption{F-measures according to the different autosimilarity matrices and kernels (SALAMI test subset, defined in~\cite{grill2015cnn}).}
 \vspace{-5pt}
\label{fig:BTF_bands_salami}
\end{figure*}

\begin{figure*}[t!]
\centering
\begin{subfigure}{0.9\columnwidth}
     \vspace{-2pt}
  \includegraphics[width=\columnwidth]{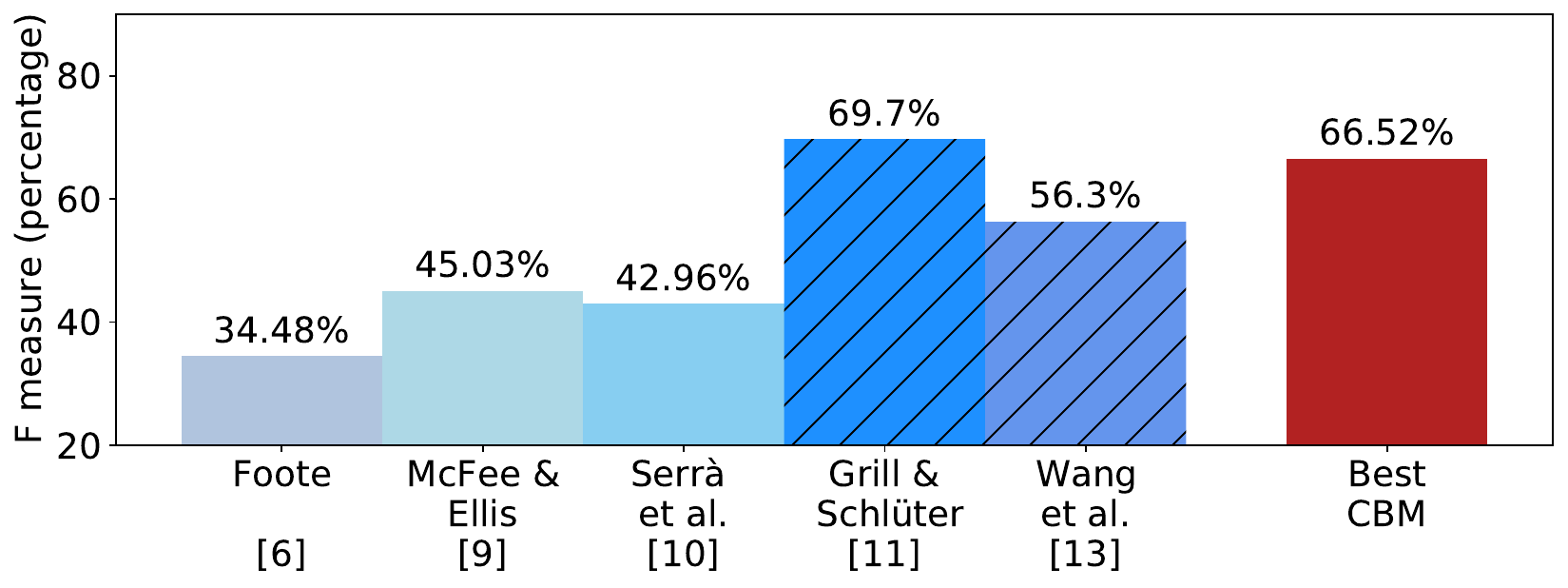}
  
  \caption{$\Fzf$.}
\end{subfigure}
\quad
\begin{subfigure}{0.9\columnwidth}
     \vspace{-2pt}
  \includegraphics[width=\columnwidth]{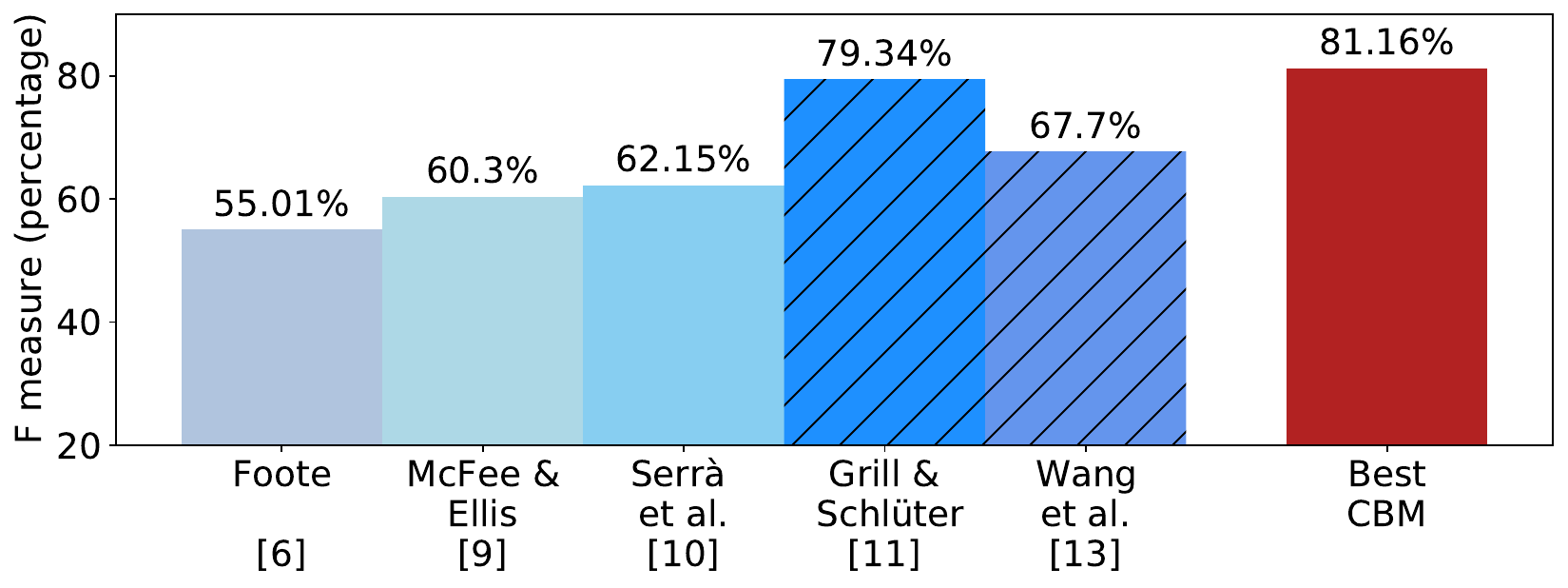}
  \caption{$\Fth$.}
  \end{subfigure}
    \vspace{-8pt}
  \caption{Boundary retrieval performance of the CBM algorithm (with the 7-band kernel) on the RWC Pop dataset, compared to State-of-the-Art algorithms. Supervised algorithms are presented with hatched bars.}
      \vspace{-5pt}
\label{fig:comparison_SOTA_rwc}
\end{figure*}

\begin{figure*}[t!]
\centering
\begin{subfigure}{0.9\columnwidth}

  \includegraphics[width=\columnwidth]{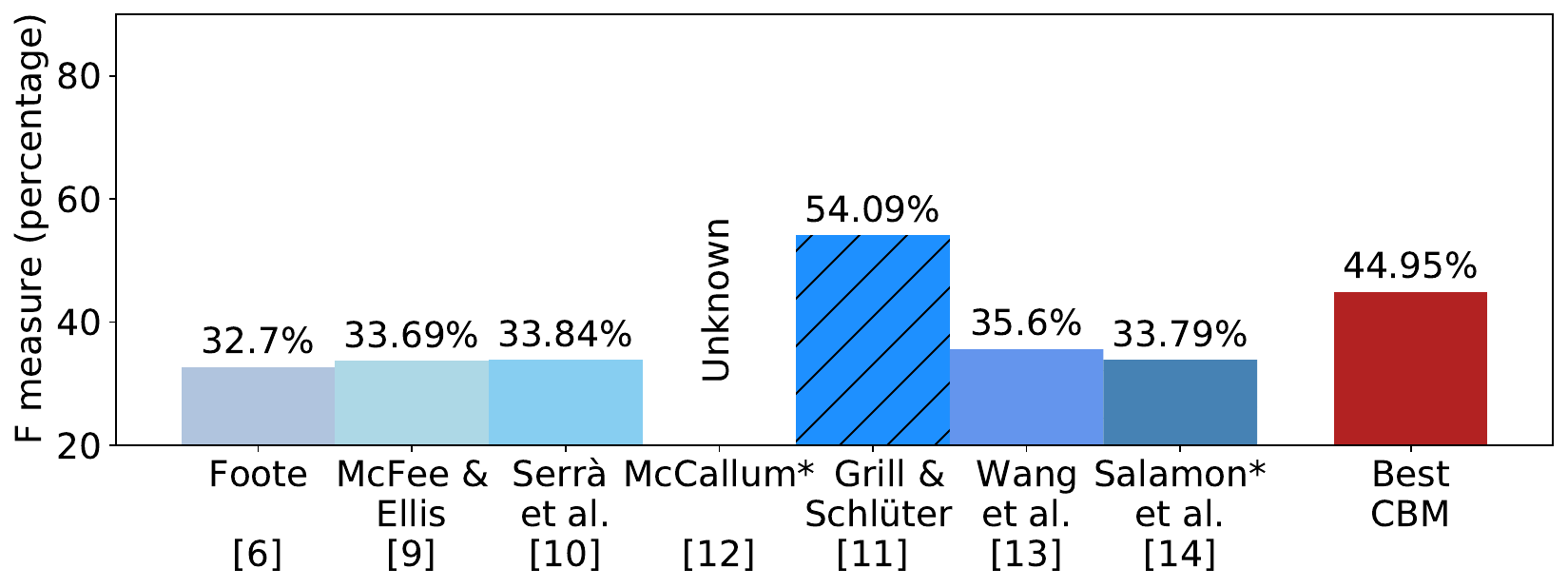}
  \caption{$\Fzf$.}
\end{subfigure}
\quad
\begin{subfigure}{0.9\columnwidth}
  \includegraphics[width=\columnwidth]{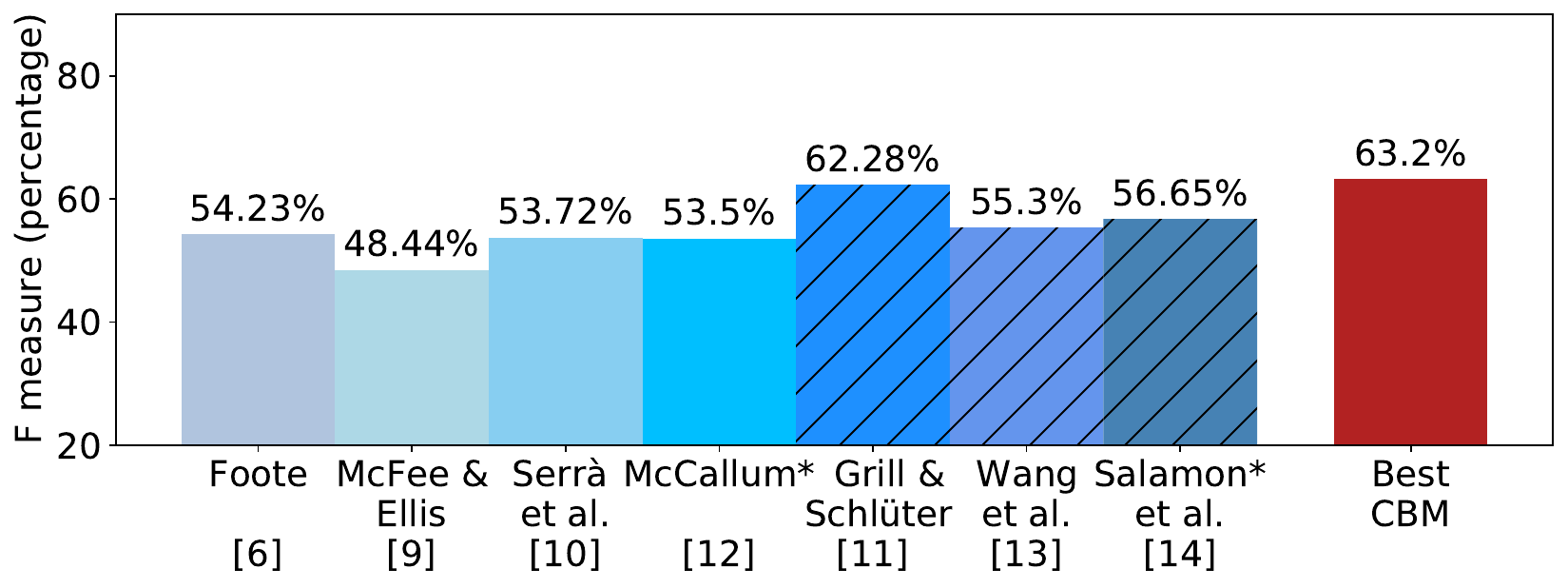}
  \caption{$\Fth$.}
  \end{subfigure}
    \vspace{-8pt}
  \caption{Boundary retrieval performance of the CBM algorithm (with the 15-band kernel) on the SALAMI dataset, compared to State-of-the-Art algorithms. Supervised algorithms are presented with hatched bars. The star * represents algorithms where the evaluation subset is not exactly the same as ours, thus preventing accurate comparison.}
      \vspace{-8pt}
\label{fig:comparison_SOTA_salami}
\end{figure*}

The CBM algorithm being a segmentation algorithm, it is evaluated on the structural segmentation task on both RWC Pop and SALAMI datasets~\cite{rwc, salami}. The SALAMI dataset is restricted to the test subset defined in~\cite{grill2015cnn}. The remainder of the dataset is used to fix parameter $\lambda$, in a learning scheme. On the RWC Pop dataset, $\lambda$ is fitted in a 2-fold cross-validation scheme, by splitting the dataset in two subsets: songs with odd \textit{vs.} even numbers, as in~\cite{marmoret2020uncovering}. For both datasets, parameter $\lambda$ takes its values between $0.1$ and $2$, with a step of $0.1$.

We evaluate the algorithm on the Hit-Rate metrics, comparing a set of estimated boundaries with a set of annotations. 
In particular, an estimated boundary $\aBound^e_i \in \BoundSet^e$ is considered to be correct if it is close (with respect to a tolerance $t$) to an annotated boundary $\aBound^a_j \in \BoundSet^a$, \ie $|\aBound^e_i - \aBound^a_j| \leq t$. Tolerances are set to 0.5s and 3s, following the standards in MSA~\cite{nieto2020segmentationreview}. The segmentation is finally evaluated with Precision, Recall and F-measure, computed using the \textit{mir\_eval} toolbox~\cite{raffel2014mireval}. Only F-measures are shown here ($\Fzf$ and $\Fth$).

The autosimilarity matrices are computed with the three similarity functions (Cosine, Covariance and RBF), and computed with the full, 3-band, 7-band and 15-band kernels. Segmentation results are presented in Fig.~\ref{fig:BTF_bands_rwc} and Fig.~\ref{fig:BTF_bands_salami} for the RWC Pop and SALAMI datasets respectively. 

These results exhibit a clear advantage of using the Covariance and RBF similarity functions compared to the Cosine similarity function. The best results on both datasets are obtained with the RBF similarity function. The design of the kernel is also largely impacting the segmentation results. In these experiments, the 7-band and the 15-band are respectively the best-performing kernels on the RWC Pop and SALAMI datasets.

Fig.~\ref{fig:comparison_SOTA_rwc} and Fig.~\ref{fig:comparison_SOTA_salami} compare the best results obtained with the CBM algorithm with those of the State-of-the-Art algorithms. In this comparison, the CBM algorithm largely outperforms the other unsupervised segmentation methods, most of the supervised algorithms, and is competitive with the global (supervised) State-of-the-Art~\cite{grill2015cnn}. These results are promising and show the potential of the CBM algorithm, which is performing well despite its relative simplicity.

All State-of-the-Art algorithms use beat-aligned features, except~\cite{grill2015cnn} which uses a fixed hop length and~\cite{wang2021supervised} which uses downbeat-aligned features. In details, results for~\cite{foote2000automatic, mcfee2014analyzing, serra2014unsupervised} are computed with the \textit{MSAF} toolbox~\cite{msaf}, and realigned on downbeats in post-processing. Results for the CNN~\cite{grill2015cnn} are extracted from the 2015 MIREX contest. Results for~\cite{mccallum2019unsupervised, wang2021supervised, salamon2021deep} are obtained from the articles themselves.

\section{Conclusions}
This article has presented different autosimilarity matrices, computed at the barscale, hence studying several distinctive ways to represent similarities between pairs of bars in a song. These autosimilarities are then processed by the CBM algorithm, estimating boundaries between structural sections.

The CBM algorithm achieves levels of performance outperforming the current unsupervised State-of-the-Art~\cite{foote2000automatic, mcfee2014analyzing, serra2014unsupervised, mccallum2019unsupervised} and comparable to those of the global (supervised) State-of-the-Art~\cite{grill2015cnn}. 

Overall, the design of the kernel strongly impacts the segmentation results. Hence, future work could focus on studying alternative types of kernels, for example using normalized values in the kernel (as in~\cite{shiu2006similarity}) and normalizing the score associated with each kernel by the number of nonzero values instead of the size of the kernel. 

Convolution kernels studied in this article focus on the homogeneity of each segment, but different kernels could be considered in order to account for the repetition criterion, \eg those of Shiu \etal~\cite{shiu2006similarity}. The kernel values could depend on the particular song or dataset considered. Of particular interest could be the learning of such kernels instead of an (empirical) definition.

Penalty values for the different cases were set quite empirically, and would benefit from further investigations.

This work opens new paths towards future progress in structural segmentation methods, for which we hope that the open-source \textit{as\_seg} toolbox~\cite{asSeg} provided with this work\footnote{\href{https://gitlab.inria.fr/amarmore/autosimilarity_segmentation/-/tree/WASPAA23}{https://gitlab.inria.fr/amarmore/autosimilarity\_segmentation/-/tree/WASPAA23}} will contribute.
\vfill

\bibliographystyle{IEEEtran}
\bibliography{refs23}
%
%
%
%
%
%
%
%
%

\end{sloppy}
\end{document}